# Acoustic properties of glubam and SPF

Jie Wen, Youbohong Kong, Yan Xiao[*], Anxu Huang, Lejun Zhou

**Abstract**

Noise pollution has attracted a wide attention due to its negative effect on human health, and many researchers are working on proposing strategies or developing new materials to improve acoustic performance of structures and building materials. Meanwhile, the worldwide emphasis on sustainable development has motivated the development and application of bio-mass materials in every field, including the engineering construction. The purpose of this study is to measure and investigate the performance of a structure-used glued laminated bamboo (Glubam) and spruce-pine-fir (SPF) as soundproof or sound-absorbing materials. The impedance tube method was used to measure the sound reduction index and sound absorption coefficient of glubam and SPF in th lab. After doing some comparison studies, it can be found that glubam and SPF have good performance in sound insulation but behave weakly in sound absorption, and glubam exhibits similar even better aoustic properties than SPF. Moreover, a finite element model was built up using ABAQUS to predict sound insulation properties of materials. According to the numercial results of sound reduction index obtained from the finite element model, developing composite materials and double-layer panles with cavity between should be effective on improving the soundproof performance of SPF and glubam.

## Introduction

Due to the development of industrialization and traffic system, noise pollution has become one of the major environmental pollution resources, which has a negative influence on human health (Goines and Hagler 2007). Many researchers have confirmed the negative effects of noise on both human health and plant growth. For instance, Ibrahim and Richard (2000) measured the noise level at schools located in residentical areas, ang they determined the negative effect of noise on teachers' and students' performance. Moreover, exposures of noise pollution have been found to be an important cause of many human health issues, including deafness, annoyance, sleep and mental disorders, heart trouble, cardiovascular disease etc (Basner et al. 2014; Miedema and Oudshoorn 2001; Muzet 2007; Sørensen et al. 2012; Stansfeld and Matheson 2003). Hence, how to reduce noise has become an important issue in the worldwide range, and increasing engineers and researchers are working on the development of strategies and materials to control the noise.

Various techniques have been proposed for the purpose of noise reduction. The benefits of applying green systems in noise attenuation of buildings have been proven by Dunnet and Kingsbury (2008). Some previous studies on the acoustic performance of trees or vegetarian belts near roads showed that vegetarian can reduce noise level through reflecting, scattering, absorbing and destructively interfere with sound waves (Azkorra et al. 2015; Van Renterghem et al. 2012). Pronk et al. (2009) measured the airbone sound insulation property of a 200 mm thick panel filled with water, and found it has similar sound insulation index ($Rw$=48 dB) compared with a 150 mm thick brick wall and 100 mm thick concrete panel. Ravinder and Belachew (2014) investigated the effect of introducing a partition wall (two wooden panels with glass wool between) on the building noise control in the exihibition hall, and the authors found that it is possible to create two acoustically divisible rooms with the partition wall. Also, some different types of walls, such as autoclaved aerated concrete walls, double skin walls and so on, were applied to ensure the requirement of sound insulation of a building (Ho et al. 2016).

In terms of materials, the conventional acoustic insulation materials used in buildings are glass wool and mineral-based materials such as rock wool or glass fiber (Li et al. 2015; Papadopoulos 2005). With the growing concern on health risks caused by these industrial or mineral fibers, as well as the increasing emphasis on energy conservation, researchers have conducted a number of studies on organic nature fibers and green materials for their acoustic characteristics (Ballagh 1996; Cucharero et al. 2021; Koizumi et al. 2003; Oldham et al. 2011; Taban et al. 2020; Zulkifh et al. 2008). For instance, Ballagh (1996) found that the natural and safe wool materials with a relative thickness can be treated as a replacement of man-made mineral fibres to achieve good acoutic performance. Zulkifh et al. (2008) investigated the sound absorption coefficient and transmission loss index of a multi-layer coir fiber panel and found it had good acoustic properties that could be treated as the potential replacement of synthetic based products. Oldham et al. (2011) investigated the sound absorption coefficients of some biomass materials, such as raw cotton, flax fibre, ramie, using a combination of impedance tube and reverberation room measurements, and their possibility and effectiveness as a porous absorber was confirmed.

Under the background of emphasizing green, environment-friendly and sustainable-development principles worldwide, promoting the development of structure-used biomass materials has become particularly important. Bamboo and wood are carbon sequestration materials and biomass materials with good mechanical properties. Researchers have conducted some studies on the acoustic performance of wood-based or bamboo-based materials. For instance, Cucharero et al. (2021) measured the acoustic absorption properties of hardwood ad softwood pulp fiber foams, and the authors concluded that further processing and smaller dimensions of these fibers contributed to better acoustic absorption performance. The sound absorption coefficients of bamboo fibers and bamboo fiberboard were measured by Koizumi et al. (2003), and the results showed bamboo fibers had better acoustic performance than plywood with the same density.

For some countries with scarcity timber resources but rich bamboo resources, like China, promoting the application of bamboo-based materials or structures have environmental and economic benefits. Since bamboo is a type of carbon sequestration material and biomass material with good mechanical properties, the scientific and reasonable promotion of bamboo's structural application will contribute to the energy conservation and emission reduction. In other words, promoting the development of bamboo-based materials and structures will help China to achieve the goal of carbon peak and carbon neutrality. Based on the above-mentioned research background, Xiao et al. (2008) developed a new type of glued laminated bamboo with a new technical term of global inspired by glued laminated timber. As a type of bamboo-based glulam, glubam can be categorized as the thin strip lamination and thick strip lamination (Xiao et al. 2010). The thin strip laminated glubam sheets are made by bidirectionally laminating approximately 2 mm thick 20mm bamboo strip mats with a longitudinal-to-transverse ratio larger than 4:1, and their typical thickness is about 10 to 30 mm (Fig. 1a). The thick strip laminated glubam is made by pressure gluing a few layers of relatively thicker bamboo strips (about 5~8mm thick and 20mm wide) (Fig. 1b). The glubam sheets are typically manufactured using a hot-pressing procedure with a typical plate size of about 2000~2500 mm long and 1200 mm wide, which can be stacked for easy transportation. Based on different design needs, the glubam sheets can be assembled by a cold-forming process, similar to wood-based glulam (Xiao et al. 2013). Some studies on glubam members have been conducted during the last decade (Li and Xiao 2016; Li et al. 2021; Wang et al. 2017; Xiao et al. 2017; Xiao et al. 2013; Yang et al. 2014), and all the results show glubam has good mechanical properties.

According to authors' literature research, studies on the sound insulation properties of wood-based and bamboo-based materials are limited, and the physical properties of the glubam in sound insulation and sound absorption have not been studied yet. Hnece, the current study aims to evaluate the acoustic properties of two types of glubam (i.e. thin strip glubam and thick strip glubam), and those of spruce-pine-fir (SPF) are also measured for comparison. All the measurements are conducted in the lab using an impedance-tube method. Some empirical and theoretical formulas are used to verify the retionality and realibility of experimental results. Moreover, a finite element model is developed in ABAQUS for the prediction of material sound reduction index, which is applied to explore some strategies and new materials to improve the sound insulation performance of glubam.

**Experimental program**

**Test equipment and theory**

In this study, a two-microphone impedance tube (Fig. 1) referring to GB/T 18696.2—2002 (GB/T 2002) and ISO 10534-2 (ISO 1998)was used to measure the acoustic absorption coefficients of glubam and SPF. The principle of the two-microphone method is: a plane sound wave is generated by the sound source, and the sound pressures at locations of two microphones A and B are recorded to determine the

transfer functions of incident wave ($H_I$), reflected wave ($H_R$) and total acoustic field ($H_{12}$) as

$$H_I = \frac{p_{1I}}{p_{2I}} = e^{-jk_0(x_1-x_2)} \tag{1}$$

$$H_R = \frac{p_{1R}}{p_{2R}} = e^{jk_0(x_1-x_2)} \tag{2}$$

$$H_{12} = \frac{p_2}{p_1} \tag{3}$$

where, $p_1$ and $p_2$ are the sound pressures measured at microphone A and B, respectively. Then, the normal incidence reflection factor $r$ and the normal incidence sound absorption coefficient $\alpha$ can be calculated using the following formulas

$$r = \frac{H_{12} - H_I}{H_R - H_{12}} e^{2jk_0 x_1} \tag{4}$$

$$\alpha = 1 - |r|^2 \tag{5}$$

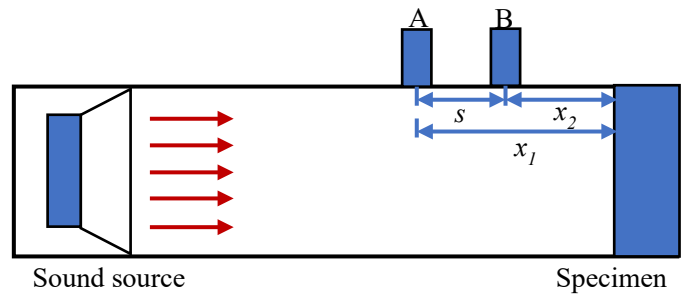

(a) Schematic diagram

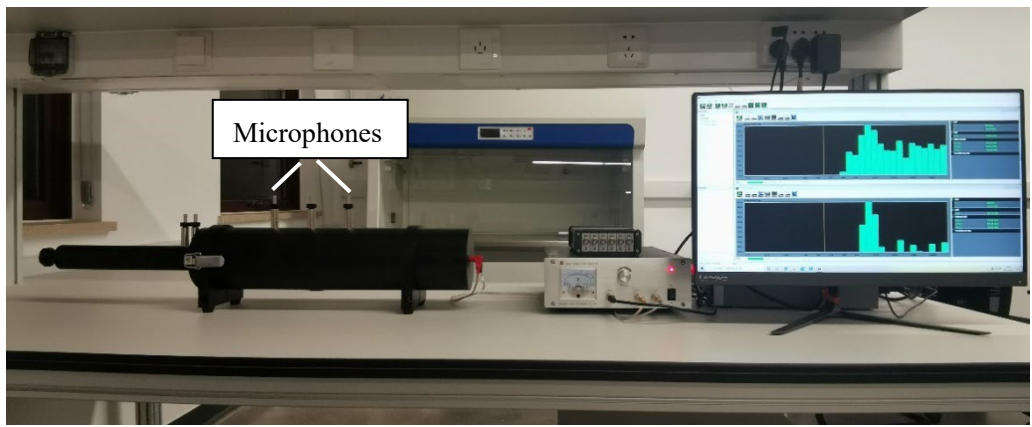

(b) Actual diagram

Fig. 1 A two-microphone impedance tube

In terms of the measurement for the sound transmission loss indexes of glubam and SPF, a four-microphone impedance tube (Fig. 2) was used according to GB/Z 27764—2011 (GB/Z 2011) and ASTM E2611 – 09 (ASTM 2009). The principle of this test can be briefly concluded as: a sound wave emanating from the sound source generates a plane sound wave in the source tube, and it will be separated in to three parts once reaching the sample surface: absorbed by the sample, reflected at the surface and penetrating through the sample; the part of the sound wave penetrating through the sample generates a plane sound wave in the receiving tube, and it will be absorbed or reflected at the absorbent ending. Two microphones are located as each side of the sample, and the standing wave separation method is applied to separate incident and reflected waves. According to the data from the four microphones, the sound reduction index $TL$ can be calculated using

$$t_p = \frac{\sin(k \cdot s_1) \cdot p_3 \cdot e^{jks_2} - p_4}{\sin(k \cdot s_2) \cdot p_1 - p_2 e^{-jks_1}} \quad (6)$$

$$TL = -20 \log_{10} |t_p| \text{ (dB)} \quad (7)$$

where, $p_1$, $p_2$, $p_3$ and $p_4$ are the sound pressures recorded by the four microphones A, B, C and D respectively, $s$ and $L$ are the distances between the sample and microphones, and $k$ is wave number.

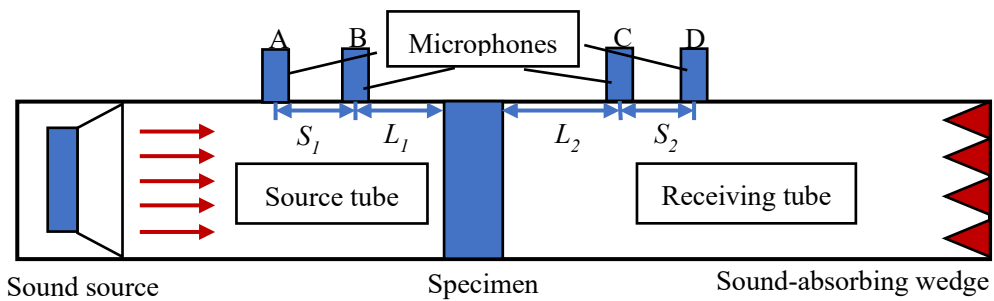

(a) Schematic diagram

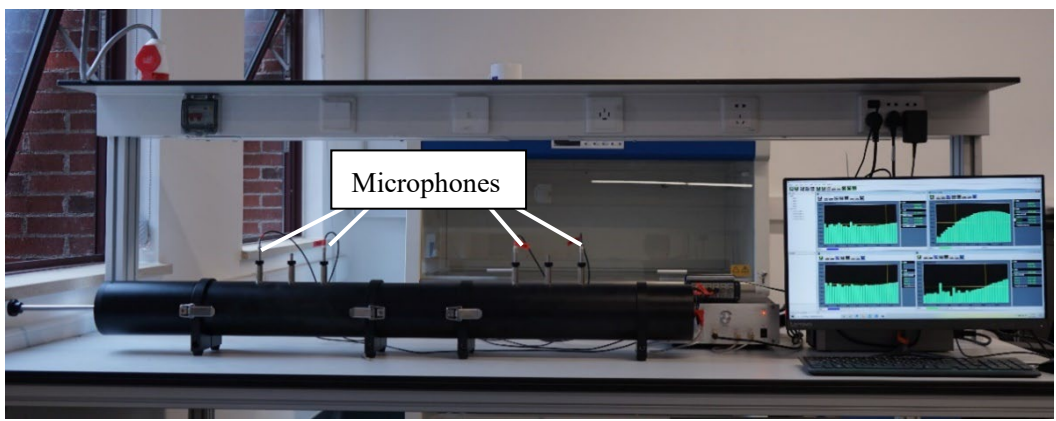

**Materials and specimens**

Thin-strip glubam, thick-strip glubam and SPF were tested in this study, and all of them are orthotropic materials. The three directions of the materials and the corresponding coordinate axis are shown in Fig. 3. Specifically, *x* axis is along the main direction of bamboo strips for glubam and the grain direction for timber, and *y* direction is perpendicular to the main bamboo-strip direction and wood-grain direction. According to the typical construction pattern of bamboo-based and wood-based materials, the acoustic property along the *z* direction of glubam and SPF plays a crucial role on the overall acoustic performance of wall or floor structures. In this study, the acoustic properties of two types of glubam and SPF in three directions were measured for comparison.

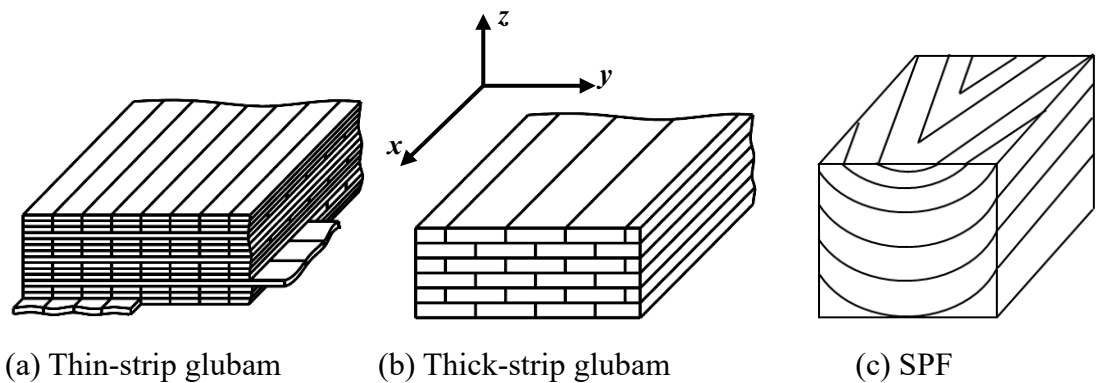

(a) Thin-strip glubam    (b) Thick-strip glubam    (c) SPF

Fig. 3 Three materials tested in this study

According to the size and introduction of the impedance tube used in this study, two-size specimens were prepared for one material to measure its sound insulation index and absorption coefficient along *z* direction in both low and high frequency range. Specifically, 29mm-diameter specimens for 500-6400Hz measurement and 100mm-diamter specimens for 50-1600Hz measurement. While, for the measurement of acoustic properties of glubam and SPF along x and y directions, only higher frequency range with 29-mm diameter specimens were considered due to the size limitation of raw materials. Two-size specimens are shown in, and three specimens were cut from one material sample in one direction for average results. The numbers and dimensions of all the specimens tested in this study are given in

Table 1 The numbers and dimensions of all the specimens

| Material | Direction | Section view | Diameter (mm) | No. | Thickness (mm) |
|---|---|---|---|---|---|
| Thin-strip glubam | $x$ | 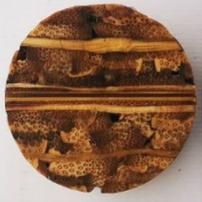 | 29 | Thin-x-1 | 28.36 |
| | | | | Thin-x-2 | 28.48 |
| | | | | Thin-x-3 | 28.99 |
| | $y$ | 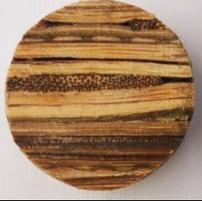 | 29 | Thin-y-1 | 28.41 |
| | | | | Thin-y-2 | 28.79 |
| | | | | Thin-y-3 | 28.64 |
| | $z$ | 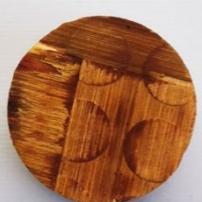 | 29 | Thin-z-H1 | 27.55 |
| | | | | Thin-z-H2 | 27.47 |
| | | | | Thin-z-H3 | 27.50 |
| | | | 100 | Thin-z-L1 | 27.66 |
| | | | | Thin-z-L2 | 27.56 |
| | | | | Thin-z-L3 | 27.17 |
| Thick-strip glubam | $x$ | 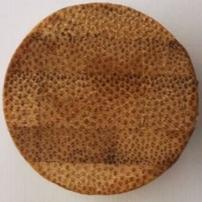 | 29 | Thick-x-1 | 31.04 |
| | | | | Thick-x-2 | 31.03 |
| | | | | Thick-x-3 | 31.60 |
| | $y$ | 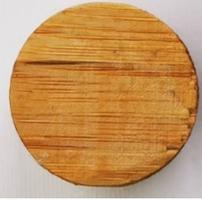 | 29 | Thick-y-1 | 31.84 |
| | | | | Thick-y-2 | 31.90 |
| | | | | Thick-y-3 | 31.87 |
| | $z$ | 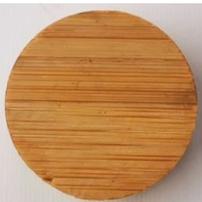 | 29 | Thick-z-H1 | 30.68 |
| | | | | Thick-z-H2 | 30.71 |
| | | | | Thick-z-H3 | 30.30 |
| | | | 100 | Thick-z-L1 | 30.58 |
| | | | | Thick-z-L2 | 31.02 |
| | | | | Thick-z-L3 | 30.65 |
| SPF | $x$ | 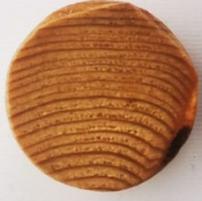 | 29 | SPF-x-1 | 35.41 |
| | | | | SPF-x-2 | 35.44 |
| | | | | SPF-x-3 | 35.54 |
| | $y$ | 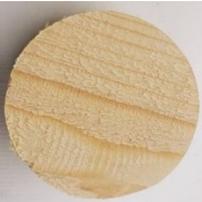 | 29 | SPF-y-1 | 34.13 |
| | | | | SPF-y-2 | 34.19 |
| | | | | SPF-y-3 | 34.24 |

| | | 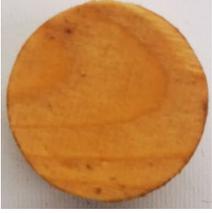 | 29 | SPF-z-H1 | 35.48 |
| | | | | SPF-z-H2 | 32.43 |
| | z | | | SPF-z-H3 | 34.40 |
| | | | 100 | SPF-z-L1 | 32.21 |
| | | | | SPF-z-L2 | 35.06 |
| | | | | SPF-z-L3 | 33.06 |

The moisture content percent of thin-strip glubam, thick-strip glubam and SPF were measured referring to GB/T 1931-2009 (GB 2009a) and ISO 3130 (ISO 1975), and the results are given in Table 2. In addition, all the materials after drying were also measured for their sound reduction indexes and sound absorption coefficients for comparison, thereby conducting a brief investigation on the effect of moisture contents on material acoustic performance.

Table 2 The moisture content percent of glubam and SPF

| Materials | Thin-strip glubam | Thick strip glubam | SPF |
|---|---|---|---|
| Moisture content percent | 9.8 % | 8.9 % | 11.8 % |

**Experimental results and discussion**

The typical transmission loss-frequency curve of a single layer material (Norton and Karczub 2007) is illustrated in Fig. 4, in where the resonance effect and coincidence effect both lead to a significant decrease of transmission reduction index.

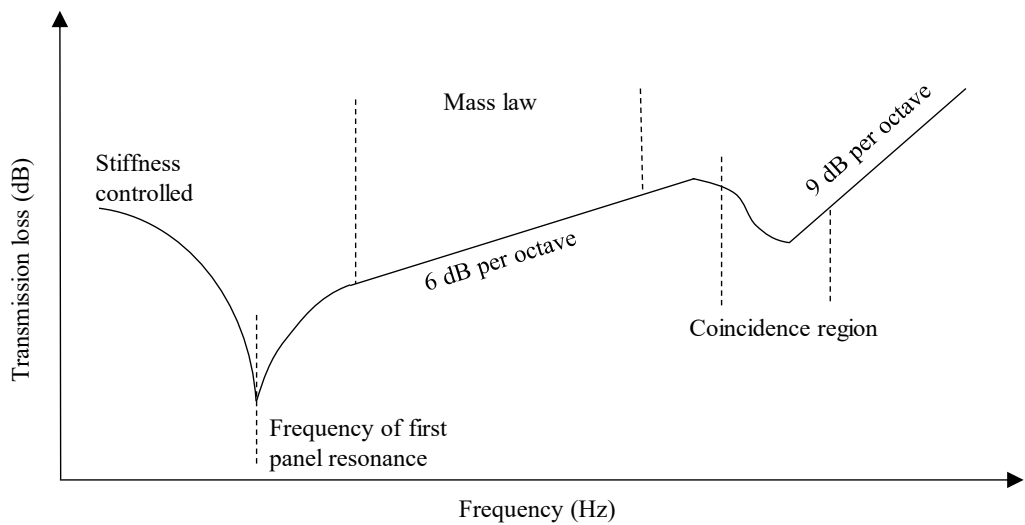

Fig. 4 Typical transmission loss-frequency curve of a single layer material

**Acoustic properties of glubam and SPF in three directions**

Both the sound reduction indexes $TL$ and absorption coefficients $\alpha$ of two types of glubam and SPF in three directions were measured in this study, and all the results are shown in Fig. 5 for comparison. As mentioned in the last section, results of $TL$ and $\alpha$ in the frequency range from 500Hz to 6400Hz were considered in this part due to the size limitation of raw materials, which were measured using 29mm-diamter specimens. For thin-strip glubam, it has similar sound insulation property in direction *y* and *z*, and there is no obvious difference between the results of sound reduction index in these two directions. While, thin-strip glubam has far weaker acoustic insulation performance in direction *x* with the lowest values of sound reduction index, which is probably caused by the visible distributed small holes in the *yoz* plane along the direction *x* (Fig. 5). Moreover, it can be found from the figure that the thin-strip glubam experiences a significant drop of sound reduction index in all the three directions around 2000Hz, which is probably due to the effect of coincidence.

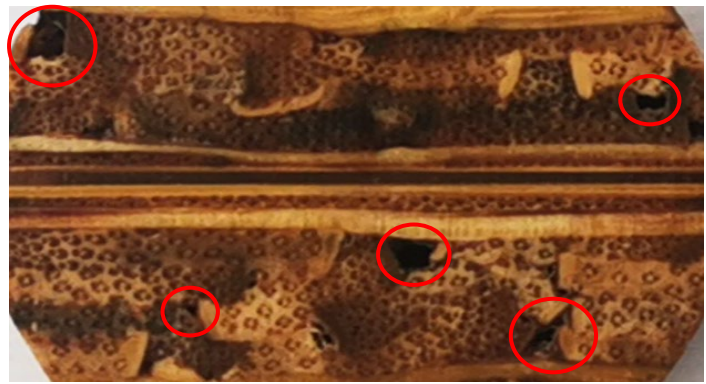

Fig. 5 *yoz* plane of thin-strip glubam with distributed visible holes

With respect to sound absorption coefficient, thin-strip glubam reveals an opposite trend that this coefficient around the first resonance frequency of thin-strip glubam achieves the top value in three directions. That is, the effect of resonance results in the poorest sound insulation performance but strongest sound absorption property. In terms of the influence of direction, the sound absorption coefficient of thin-strip glubam also has the different situation than the sound reduction index. Specifically, thin-strip glubam has the best sound absorbing performance along the x direction, which is similar along *y* and *z* directions at the range from about 1000Hz to 6400Hz. However, from 500Hz to 1000Hz, the sound absorbing coefficient of thin-strip glubam in *y* direction is found to be significantly higher than that in *z* direction.

The results of acoustic properties of thick-glubam and SPF (see the Fig. 5(b) and (c), respectively) show a similar relationship between acoustic insulation and absorption performance to that observed for thin-strip glubam, that is, a larger sound reduction index is usually accompanied by a lower sound absorption coefficient and vice versa. Specifically, thick-strip glubam has similar sound reduction indexes in direction *y* and *z* but a bit lower value in direction *x*; correspondingly, its largest sound absorption coefficient happens in direction *x* with similar and lower values in directions *y* and *z*. For SPF, it has the lowest sound insulation but highest sound absorption properties in direction *x*, which is probably due to the structure of the longitudinal tracheid for coniferous wood along the tree-trunk direction *x* (Fig. 6). The sound reduction indexes of SPF in *y* direction are a bit higher than those in *x* direction, while its sound absorption coefficient in *y* direction is a bit lower than that in *x* direction.

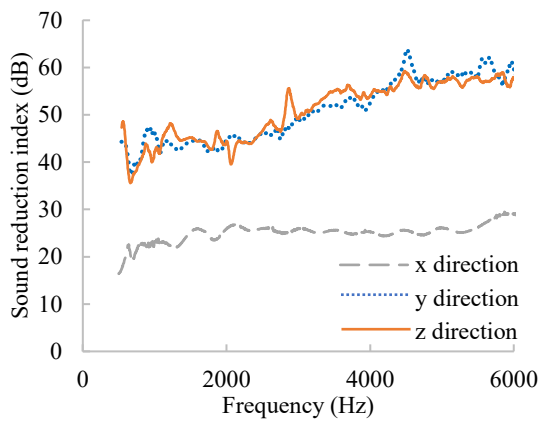 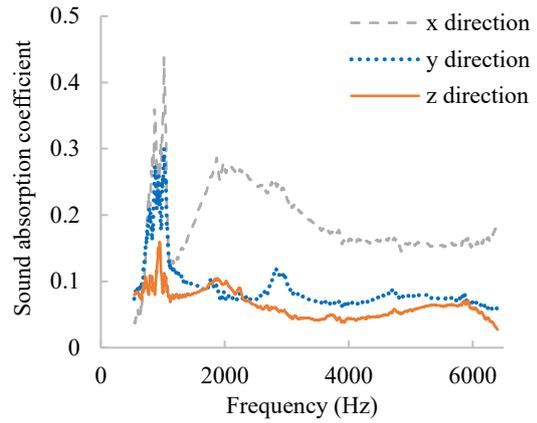

a1. Sound reduction index      a2. Sound absorption coefficient

(a) Thin-strip glubam

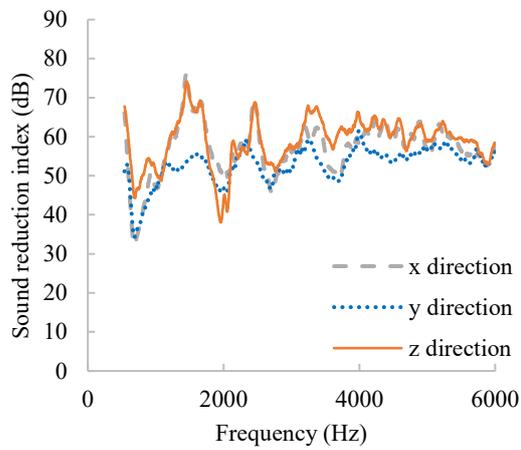 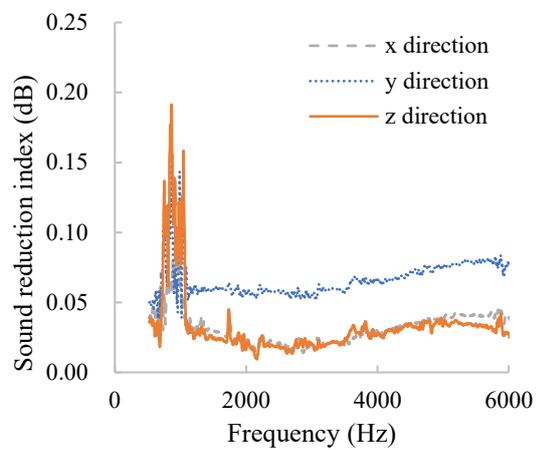

b1. Sound reduction index      b2. Sound absorption coefficient

(b) Thick-strip glubam

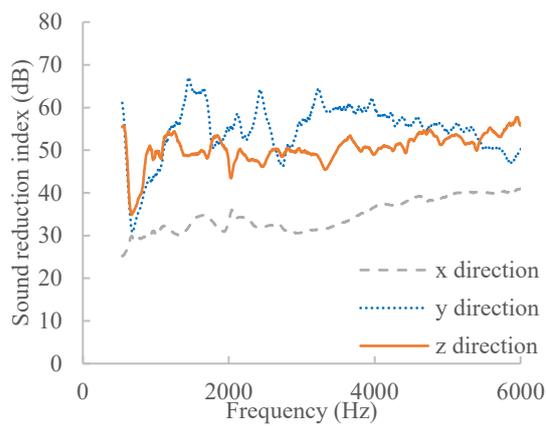 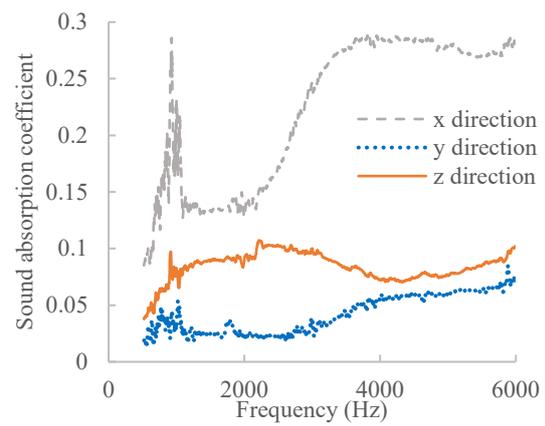

c1. Sound reduction index      c2. Sound absorption coefficient

(c) SPF

Fig. 6 Sound reduction indexes and absorption coefficients of glubam and SPF in three directions

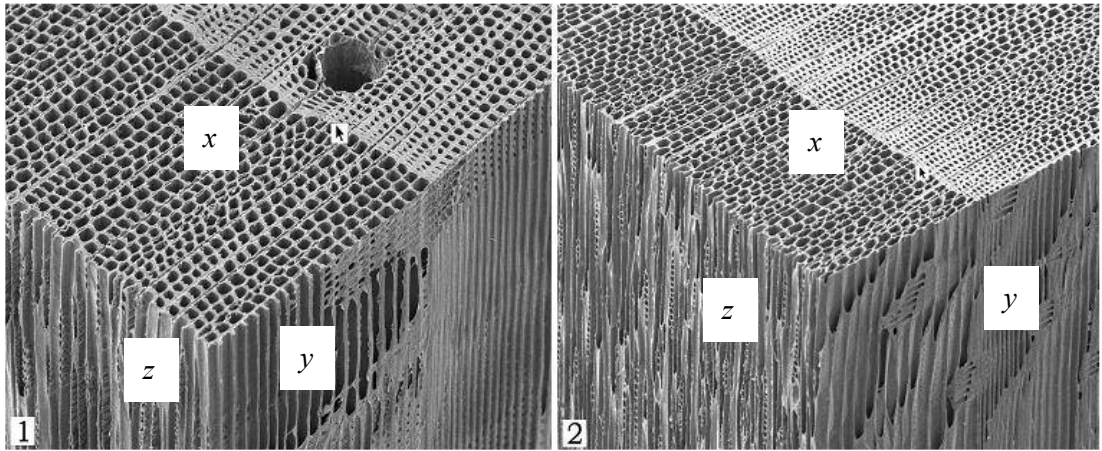

Fig. 7 Three-dimensional microscopic structure of coniferous wood

**Effect of moisture content**

As mentioned before, two moisture conditions (i.e., absolutely dry and natural environment) were considered in this acoustic experiment to investigate the effect of moisture content percent. The results of both sound reduction index and sound absorption coefficient of thick-strip glubam with two different values of moisture content percent are plotted in Fig. 7, and those results for thin-strip glubam and SPF are shown in Fig. 8 and Fig. 9 respectively.

Fig. 7 shows that for thick-strip glubam, higher moisture content percent results in higher sound reduction index but lower sound absorption coefficient as expected, and the difference exhibits more significant at low frequency range. it should be noted that the change in moisture content percent does not cause the change of some critical frequencies for thick-strip glubam, which means that the peaks and valleys of both sound reduction index and absorption coefficient almost happened at same frequency for specimens with different moisture content percents. The moisture content percent behaves a similar influence on the acoustic performance of SPF (Fig. 9). However, the situation of thin-strip glubam is a bit different, where the effect of moisture content is not obvious on its acoustic properties (Fig. 8). This is probably due to the existence of some small visible holes in thin-strip glubam (Fig. 5).

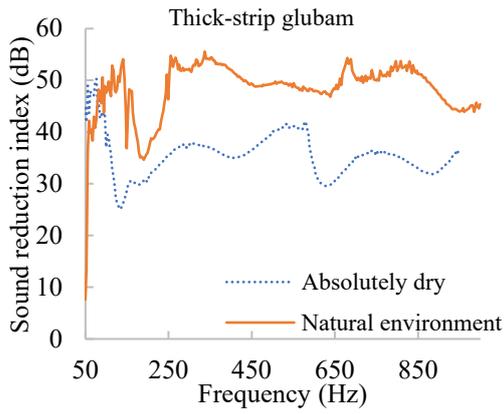
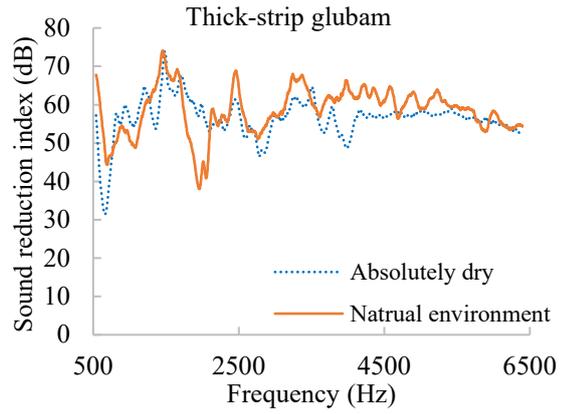

a. Sound reduction index at low frequency

b. Sound reduction index at high frequency

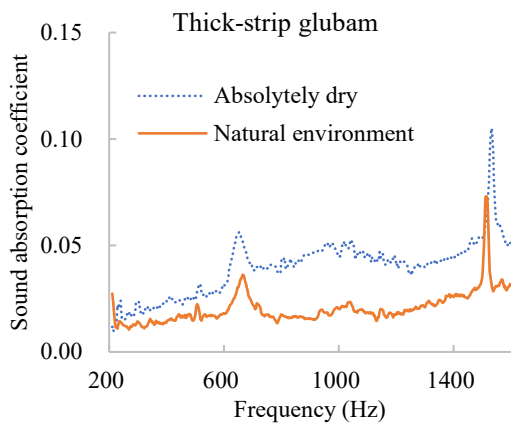
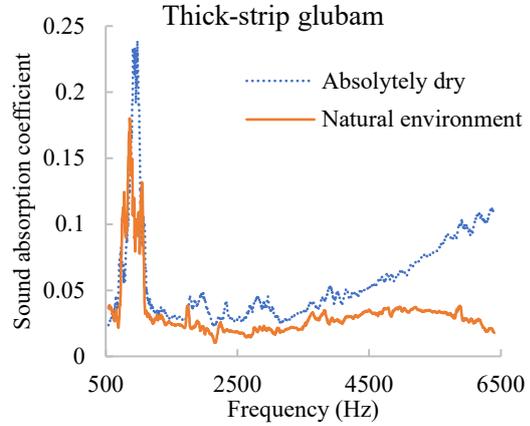

c. Sound absorption coefficient at low frequency

d. Sound absorption coefficient at high frequency

Fig. 8 Acoustic properties of thick-strip glubam

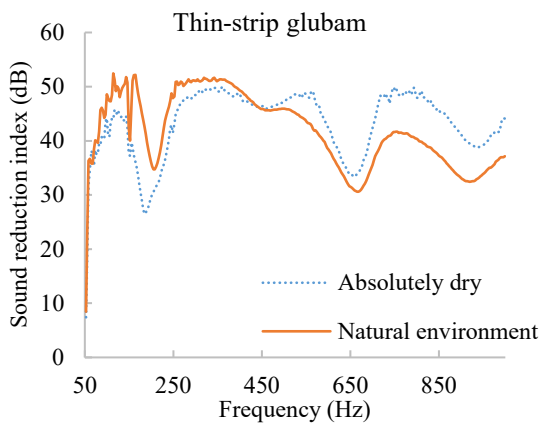
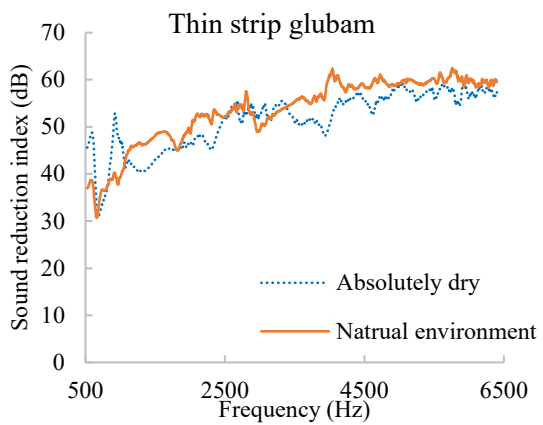

a. Sound reduction index at low frequency

b. Sound reduction index at high frequency

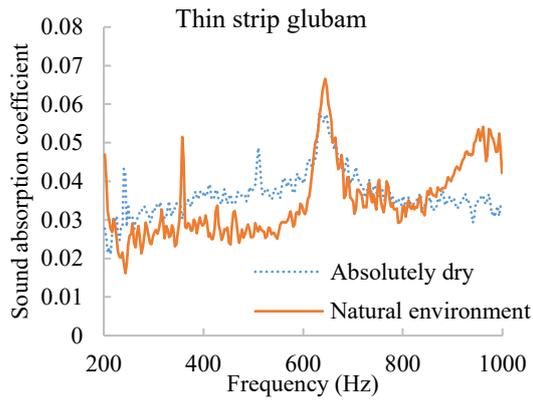
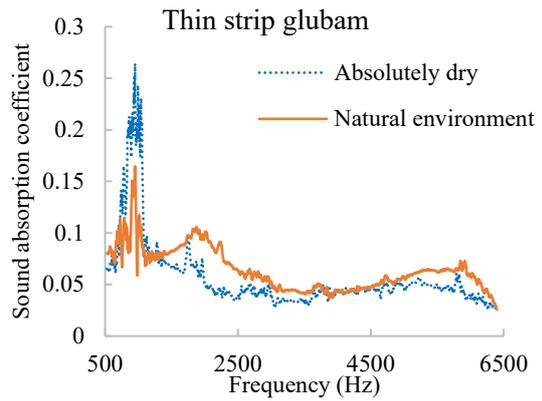

c. Sound absorption coefficient at low frequency

d. Sound absorption coefficient at high frequency

Fig. 9 Acoustic properties of thin-strip glubam

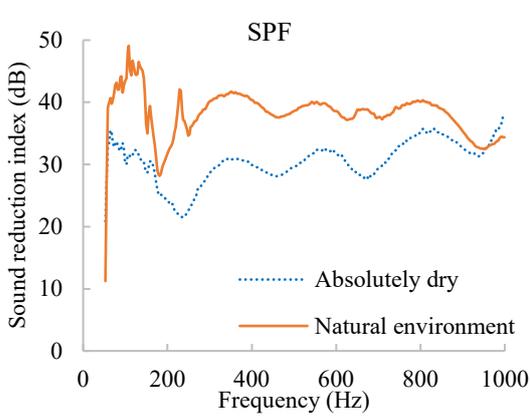
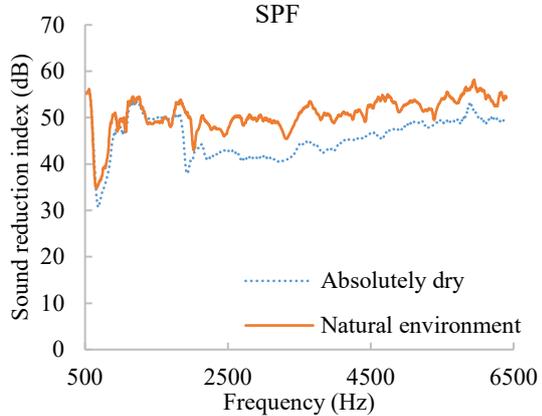

a. Sound reduction index at low frequency

b. Sound reduction index at high frequency

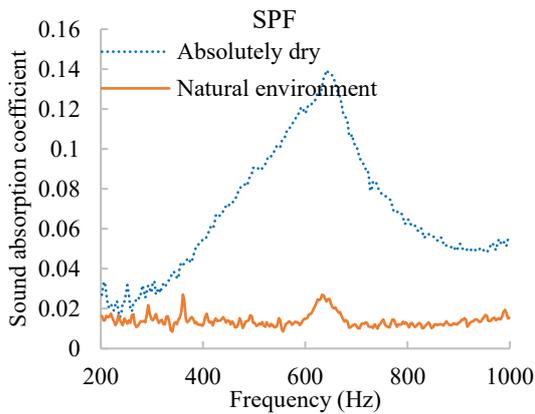
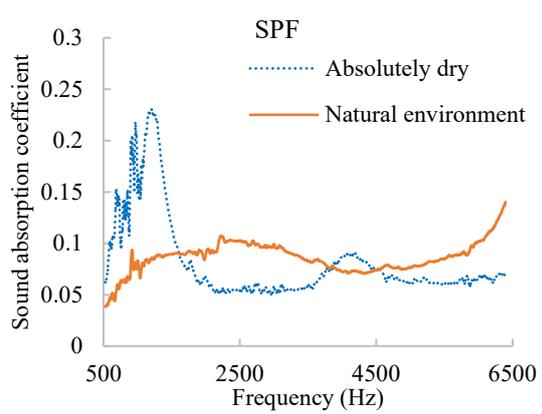

c. Sound absorption coefficient at low frequency

d. Sound absorption coefficient at high frequency

Fig. 10 Acoustic properties of SPF

**Evaluation of the acoustic performance of glubam and SPF**

Based on the application of glubam and SPF in practice, only their acoustic properties along $z$ direction are paid attention in this part for comparison, and all the results, including sound reduction index and absorption coefficient, are given in Fig. 7. It can be found that thick-strip glubam exhibits the best sound insulation performance among the three materials in both the low and high frequency ranges, while SPF has the relatively poorer behavior in sound proof. In terms of the sound absorbing performance, the situation among the three materials is different from low frequency to high frequency. In low frequency range, thin-strip glubam shows a significant advantage in sound absorption, and thick-strip glubam and SPF have similar values of sound absorption coefficient. However, in high frequency, the material with the highest absorption coefficient is SPF, which is followed by thin-strip glubam and then thick-strip glubam.

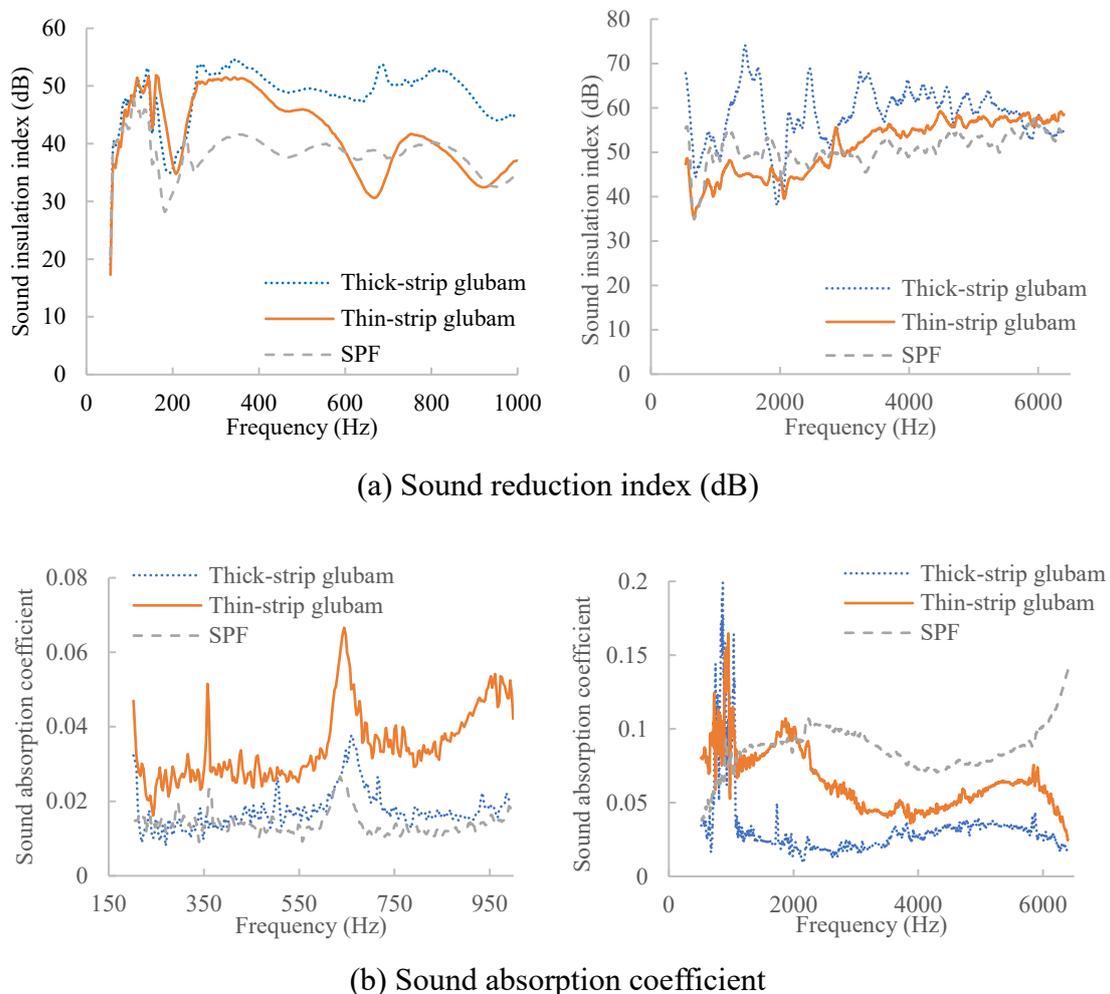

(a) Sound reduction index (dB)

(b) Sound absorption coefficient

Fig. 11 Acoustic properties of thin-strip glubam, thick-strip glubam and SPF

In order to evaluate the acoustic performance of thin-strip glubam, thick-trip glubam and SPF, the acoustic properties of some typical sound insulation and absorption materials are referred for comparative studies. Collings and Stewart (2011) measured the normal transmission loss of a 75 mm thick isotropic Autoclaved Aerated Concrete (AAC) panel in the frquency range of 50 Hz to 1600 Hz uisng an impedance tube, and Zhao et al. (2010) investigated the sound insulation property of a wood and used tire rubber composite panel based on a four-microphone method. The sound absorption coefficients of glass wool and mineral wool are given by Asdrubali et al. (2012). All the results are plotted in Fig. 8 for comparing with those of glubam and SPF measured in this study. Specifically, two types of glubam and SPF all shows good soundproof performance by comparing their sound insulation indexes with those of some sound insulation materials, such as AAC and wood-based composite material. On the contrary, thick-strip glubam, thin-strip glubam and SPF are found not adequate to be sound-absorbing materials due to their signifacantly lower sound absorption coefficients comapared with polyester fibre and glass wool. Therefore, the following parts in this study will focus on the sound insulation properties of the two types of glubam and SPF.

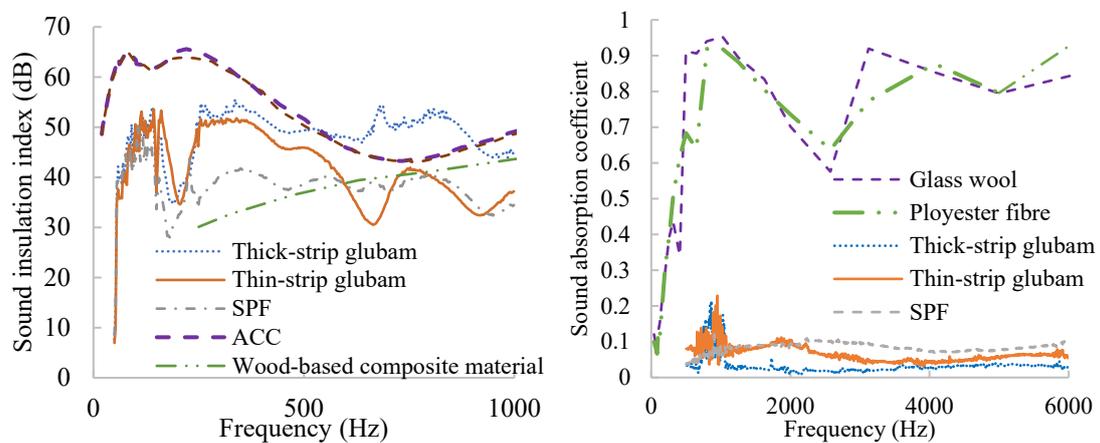

Fig. 12 Acoustic properties of glubam and SPF compared with other materials

**Prediction of Young's modulus**

According to the typical transmission loss-frequency curve given in Fig. 4, the values of the coincidence frequency of thick-strip glubam, thin-strip glubam and SPF can be directly obtained from the results of sound transmission loss measured in this study. Then, according to the following equation (Eq. 8) for the prediction of the coincidence frequency, the value of young's modulus ($E$) of each material can be predicted using Eq. 9. In order to assess the accuracy of the predicted values, the modulus of elasticity in compression along the $z$ direction for glubam and SPF was measured according to GB/T 1943-2009 (GB 2009b) with the results given in Table 3 as well as the predicted results. Table 3 shows the reasonable agreement between the measured and predicted results of young's modulus in direction $z$.

$$f_c = \frac{c^2}{2\pi}\sqrt{\frac{12M(1-v^2)}{Et^3}} \qquad (8)$$

$$E = \frac{12M(1-v^2)}{(2\pi f_c/c^2)^2 \cdot t^3} \tag{9}$$

where, $c$ is the sound speed in air, $M$ is the surface density (kg/m²), $t$ is the material thickness, $E$ is young's modulus, and $v$ is Poisson's ratio.

Table 3 The coincidence frequency $f_c$ and young's modulus of glubam and SPF along $z$ direction as well as that predicted using acoustic results

| Material | $f_c$ (Hz) | Predicted $E$ (GPa) | Measured $E$ (GPa) | Error (%) |
|---|---|---|---|---|
| Thin-strip glubam | 1813.5 | 1.280 | 1.102 | 16.18 |
| Thick-strip glubam | 1066.4 | 2.242 | 2.380 | 5.79 |
| SPF | 1696.3 | 0.550 | 0.569 | 3.32 |

**Prediction of sound reduction index**

Some different expressions and models are employed in this part to predict the sound transmission loss of two types of glubam and SPF to verify the accuracy of experimental results. The one of the most basic and well-known expressions for the prediction of sound transmission loss (*TL*) is the mass law (Callister and George 1995; Norton and Karczub 2007), which assumes that the *TL* is a function of sound frequency and surface density of the sample as expressing in Eq. 10. The agreement between the results obtained from the mass law and experiments is acceptable (Fig. 4 ); however, the mass law does not consider the coincidence effect. Then, the Sharp's model (Sharp 1978) was applied to calculate the sound transmission loss of materials to introduce the influence of coincidence effect (Eq. 11). Sharp (1978) divided the model into three parts according to the coincidence frequency, which can be calculated using Eq. 8. The calculated results of sound reduction index of glubam and SPF using Sharp's method are exhibited in Fig. 10 with experimental results, and Fig. 10 also shows these two results are in fairly good accordance. The comparison studies conducted between test results and empirical values as well as theoretical results can be used to confirm the reliability of the experimental results measured in this acoustic study.

$$TL = 10\log 10\left(1 + \left(\frac{m\pi f}{\rho c}\right)^2\right) - 5 \tag{10}$$

where, $m$ is surface density (kg/m²), $f$ is the sound frequency, $\rho$ is the density of air, and $c$ is the sound speed in the air.

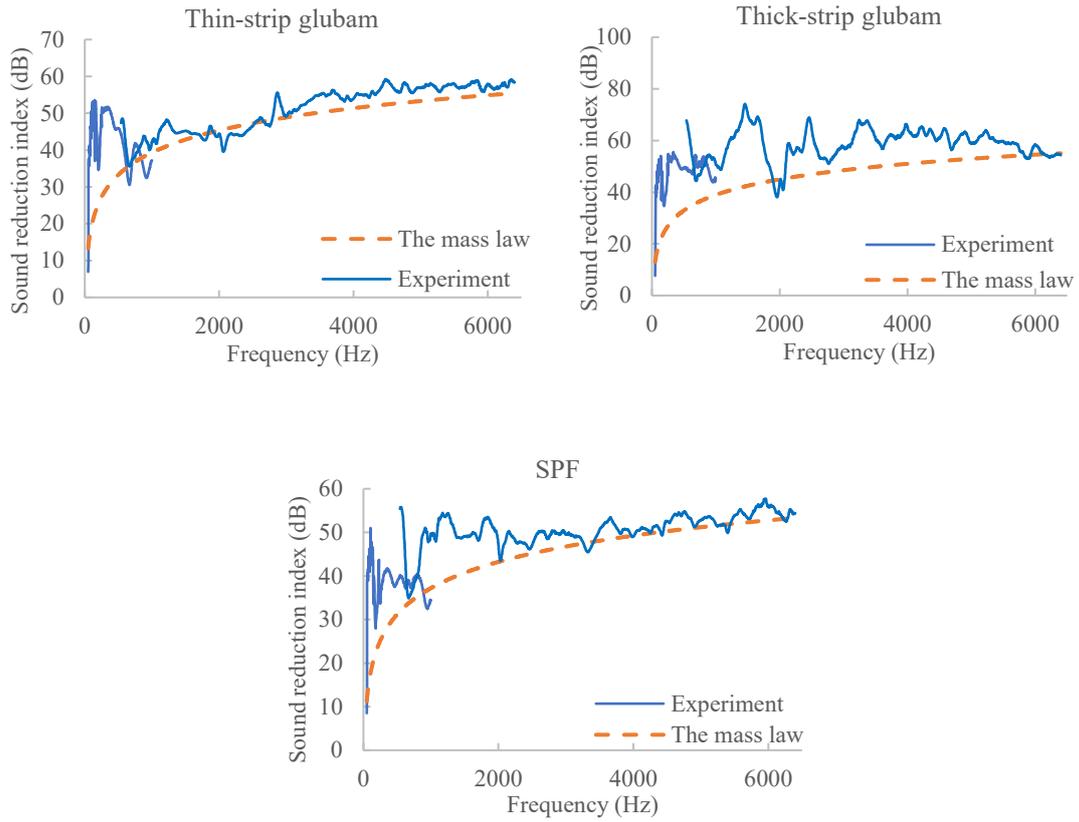

Fig. 13 Sound reduction index of glubam and SPF obtained from experiment and the mass law

$$f \leq f_c, TL = 20\log_{10}(mf) - 48$$

$$f = f_c, TL_c = 20\log_{10}(mf_c) + 10\log_{10}(\eta) - 40$$

$$f > f_c, TL = TL_c + 30\log_{10}(f/f_c)$$

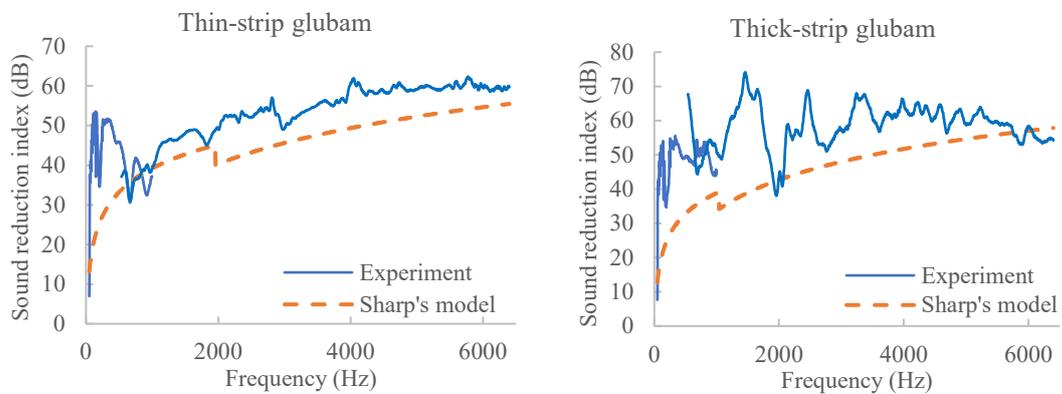

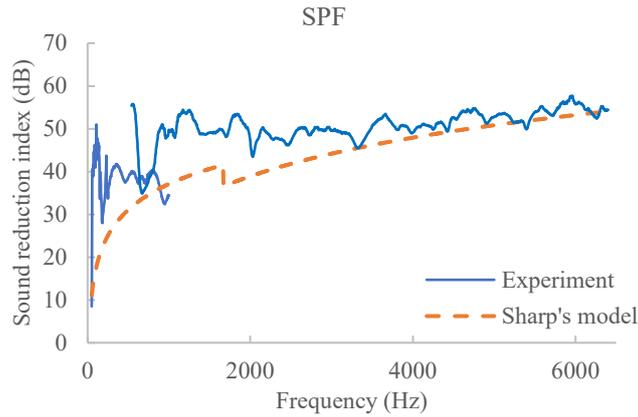

Fig. 14 Sound reduction index of glubam and SPF obtained from experiment and Sharp's model

**Finite Element Model**

A detailed finite element model for the prediction of materials sound insulation property is also built up using a well-known commercial finite element software ABAQUS. Acoustic elements AC3D20 are used to simulate a 100mm diameter impedance tube with a 480 mm long source tube and a 580 mm long receiving tube. A perpendicular incident acoustic pressure is applied at the initial end of the source tube for the simulation of plane sound wave in the impedance tube, and the nonreflecting boundary condition is set up at the end of receiving tube for the absorbing end. A three-dimensional stress element is used to the material specimen, and tie constraint is between acoustic element and solid element to take the sound-solid interaction into action. Four points are chosen in the finite element model as the positions of 4 microphones in an impedance tube (two on source tube and two on receiving tube). with the values of complex acoustic pressure at these four points, numerical results of material sound transmission loss can be calculated according to Eqs. (6) and (7).

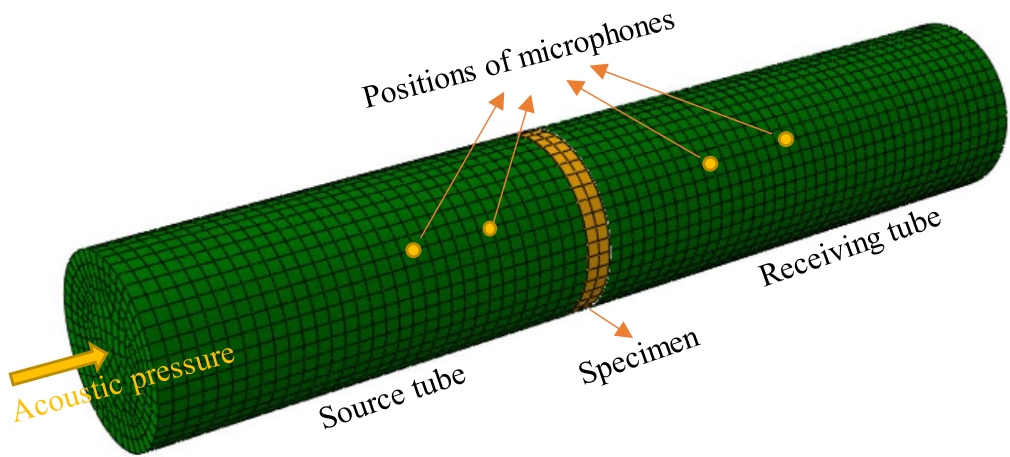

Fig. 15 Finite element model for impedance tube

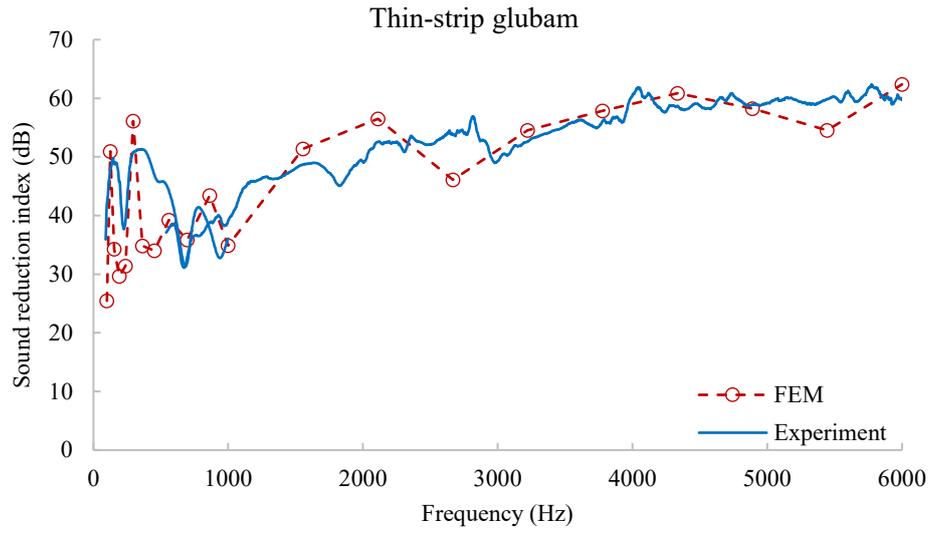

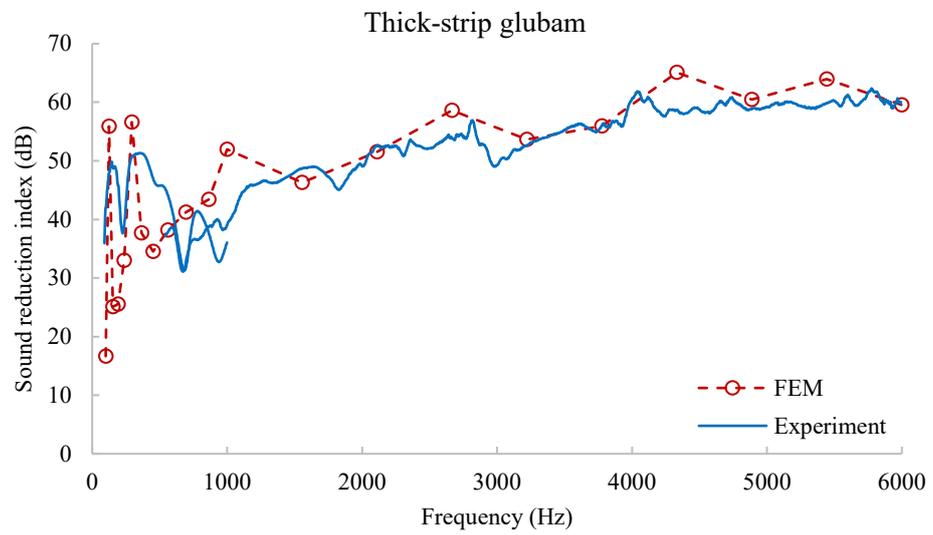

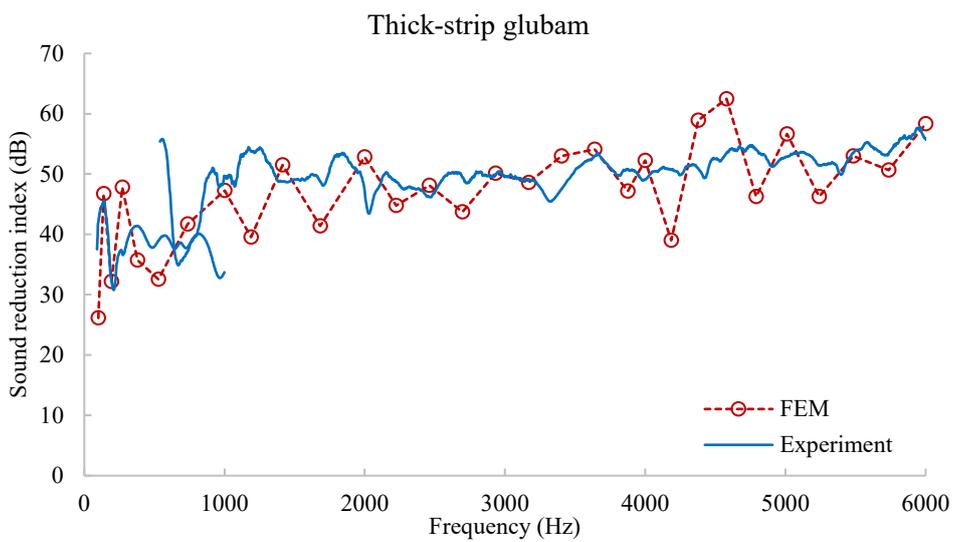

Fig. 16 Sound reduction index of glubam and SPF obtained from experiment and finite element model

The predicted results of sound reduction index for three materials along *z* direction using the finite element model are plotted in Fig. 12 with experimental results for comparison. Fig. 12 shows the good agreement between numerical and experimental results, which confirms the accuracy of the finite element model.

**Design Recommendation**

The finite element model using ABAQUS was applied to explore the effectiveness of some strategies and new materials on increasing sound insulation properties of wood-based and bamboo-based materials. Fig. 17 plotted sound reduction index (along z direction) curve related to frequency for three materials with different thickness, which indicates that the increase of specimen thickness achieves the increase of soundproof performance of thin-strip glubam, thick-strip glubam and SPF. However, the numerical results shown in Fig. 17 confirms that the improvement of materials' sound reduction index due to the increasing 10 mm thickness is not significant.

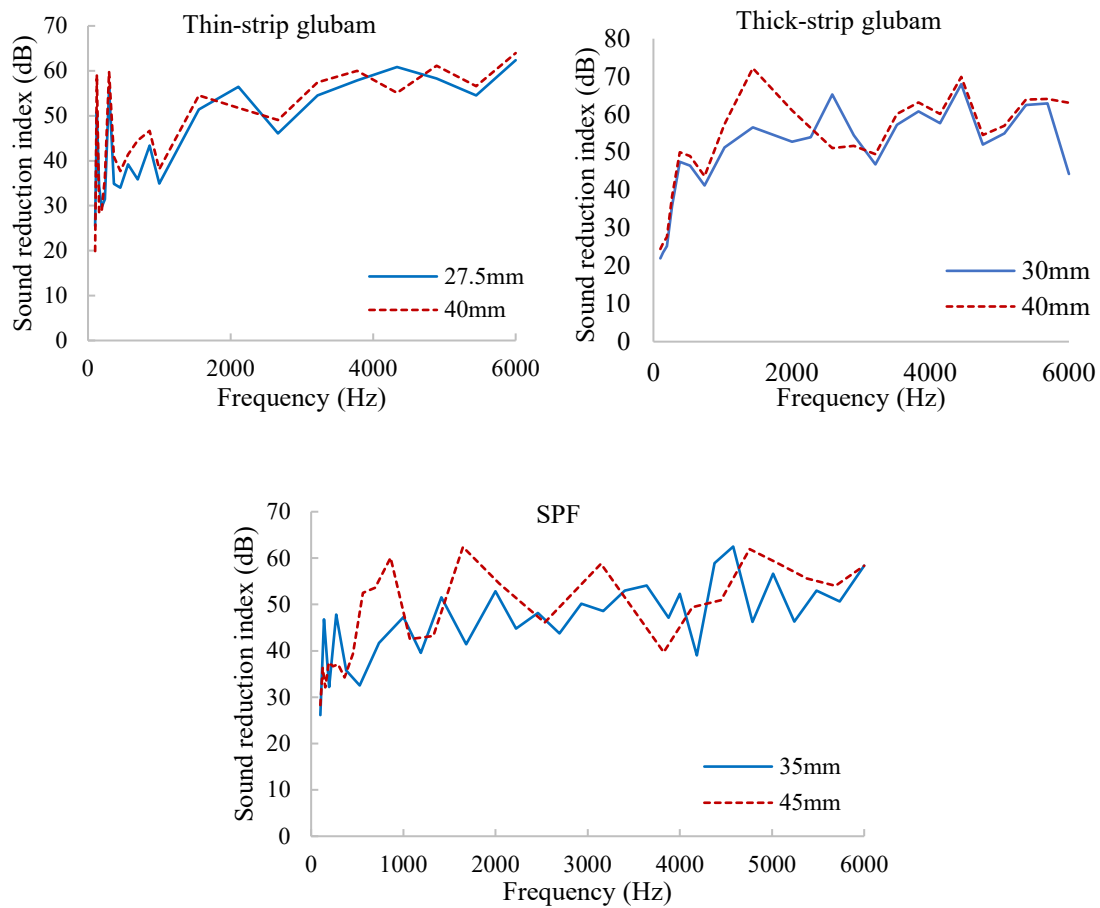

Fig. 17 Sound reduction index of three materials with different thickness along *z* direction

Then, some composite materials of glubam and SPF are proposed to obtain better sound insulation performance. Experimental results found that thick-strip glubam has the best performance in sound insulation, followed by thin-strip glubam and SPF came last (Fig. 11). Based on the finite element results (Fig. 18) obtained from the Abaqus model, a type of glubam-timber composite material composed of 2 layers of 12.5 mm thick SPF and 10 mm thin-strip glubam can achieve higher sound reduction index than SPF with same thickness. Similarly, combining thin-strip glubam with thick-strip glubam results in the increasement of sound insulation performance for thin-strip glubam (Fig. 19).

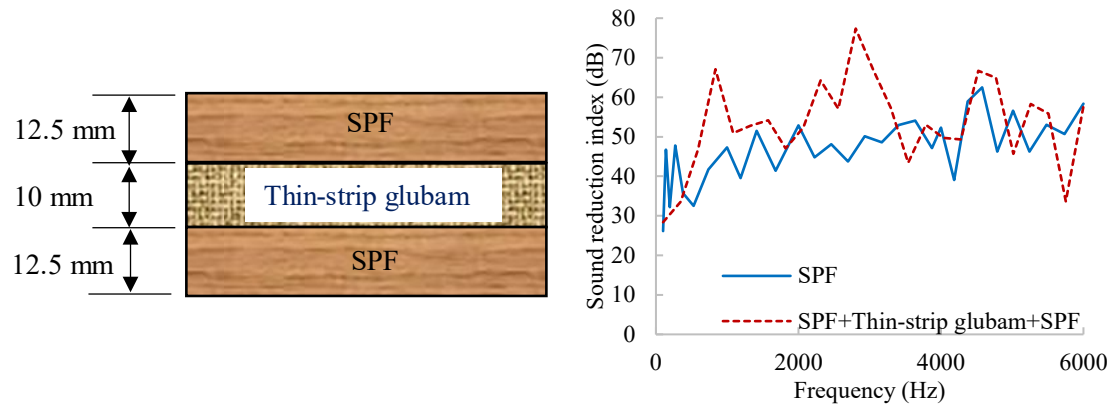

Fig. 18 Composite material made of SPF and thin-strip glubam and its sound reduction index

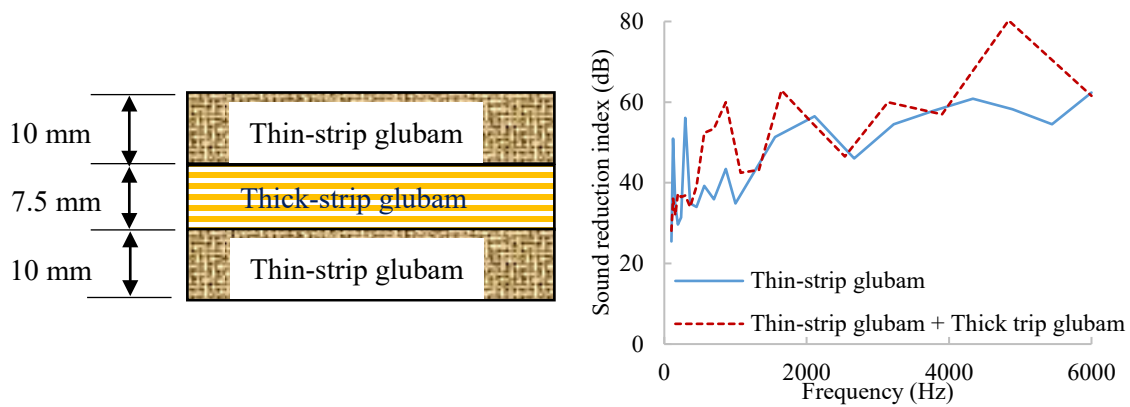

Fig. 19 Composite material made of thin-strip glubam and thick-strip glubam and its sound reduction index

Inspired by the double-layer walls' sound insulation properties, double-layer SPF panel with cavity between layers was analyzed by the proposed finite element model in ABAQUS for their soundproof performance. Fig. 20 shows that double-layer SPF achieves much better sound insulation performance than single-layer SPF as expected. The same situation also happens on thin-strip glubam and thick-strip glubam specimens, where double-layer glubam panels exhibit significantly higher sound reduction index than single-layer glubam panels (Figs. 21-22).

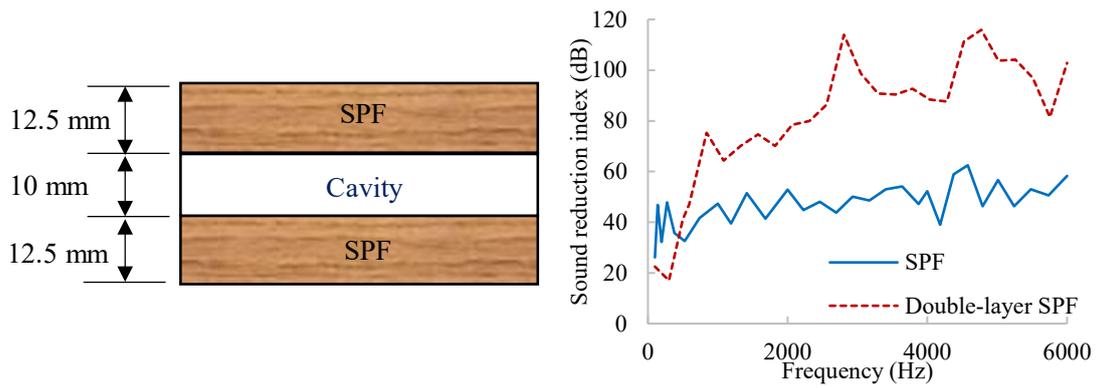

Fig. 20 Sound reduction index of double-layer SPF panel with cavity between layers

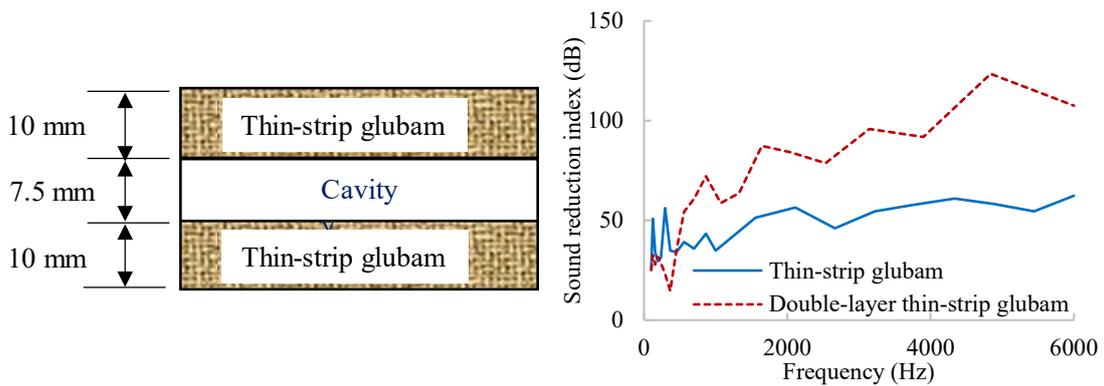

Fig. 21 Sound reduction index of double-layer thin-strip glubam panel with cavity between layers

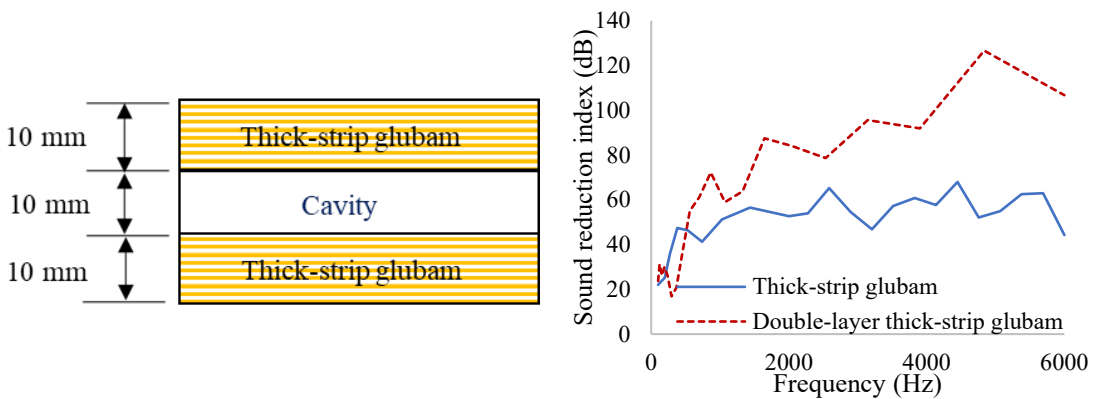

Fig. 22 Sound reduction index of double-layer thick-strip glubam panel with cavity between layers

**Conclusion**

In summary, this study investigated the acoustic performance of two types of glubam and SPF in terms of sound insulation and sound absorption. An impedance tube was used to measure the sound reduction index and sound absorption coefficient for thin-strip glubam, thick-strip glubam and SPF. After doing some comparison and analysis, all the findings in the acoustic performance of two types of glubam and SPF can be concluded as,

1, Considering the anisotropic properties of bamboo and timber, the corresponding acoustic properties along three directions were measured for all the materials in this test. The experimental results show obvious differences on both sound insulation and absorption coefficients for glubam and SPF due to their different properties in different directions.

2, Glubam and SPF specimens under absolutely dry and natural conditions were both considered to investigate the effect of moisture content on materials' acoustic properties. The effect is found to be significant for thick-strip glubam and SPF with the trend that, the increase of moisture content percent resulted in the increasing sound reduction index but decreasing sound absorption coefficient. On the other hand, different values of moisture content percent for thin-strip glubam have no obvious influence on its acoustic properties.

3, Among the tested three materials (i.e., thin-strip glubam, thick-strip glubam and SPF), thick-strip glubam exhibits the best sound insulation performance, while SPF has the relatively poorer behavior in sound proof. In terms of sound absorption, thin-strip glubam shows the best acoustic absorbing performance in low frequency range, while its advantage is replaced by SPF in high frequency range. In general, thin-strip glubam, thick-strip glubam and SPF can be used as sound insulation materials but not sound absorbents since they have relatively high sound reduction index and quite low sound absorbing coefficient.

4, The coincidence frequency of each material can be read from the sound reduction index curve obtained from the acoustic experiments, which can be used to do a quick and simple prediction of young's modulus.

5, Based on the finite element model in ABAQUS, it can be found that increasing material thickness can improve its sound insulation performance to some extent, but it is not an economy and effective way. On the other hand, developing composite materials and double-layer panels with cavity between contributes to the increasing sound reduction index for SPF, thin-strip glubam and thick-strip glubam.

**Acknowledgement**